# Highly Sensitive Fluorescent pH Microsensors Based on the Ratiometric Dye Pyranine Immobilized on Silica Microparticles

Dr. Anil Chandra[a,§,*], Dr. Saumya Prasad[a,§], Dr. Helena Iuele[a], Francesco Colella[a], Dr. Riccardo Rizzo[a], Eliana D'Amone,[a] Prof. Giuseppe Gigli[a,b] and Dr. Loretta L. del Mercato[a,*]

| | |
|---|---|
| [*] | Corresponding authors: |
| | Dr. Anil Chandra |
| | E-mail: anilchandra87@gmail.com |
| | Dr. Loretta L. del Mercato |
| | E-mail: loretta.delmercato@nanotec.cnr.it |
| [a] | Institute of Nanotechnology of National Research Council (CNR-NANOTEC), |
| | c/o Campus Ecotekne, via Monteroni, 73100, Lecce, Italy |
| [b] | Department of Mathematics and Physics "Ennio De Giorgi" University of Salento, |
| | via Arnesano, Lecce 73100, Italy |
| [§] | These authors contributed equally |
| | Supporting information for this article is given via a link at the end of the document |

**Abstract:** Pyranine (HPTS) is a remarkably interesting pH sensitive dye that has been used for plenty of applications. Its high quantum yield and extremely sensitive ratiometric fluorescence against pH change makes it a very favorable for pH sensing applications and development of pH nano/microsensors. However, its strong negative charge and lack of easily modifiable functional groups makes it difficult to be used with charged substrates such as silica. This study reports a noncovalent HPTS immobilization methodology on silica microparticles that considers the retention of pH sensitivity as well as long term stability of the pH microsensors. The study emphasizes on importance of surface charge for governing the sensitivity of the immobilized HPTS dye molecules on silica microparticles. Importance of methodology of immobilization that preserves the sensitivity as well as stability of the microsensors is also assessed.

pH sensors based on small molecular dyes have been used in conjugation with various nano/microparticles to enhance their properties.[1] As a substrate, fluorescent nano/microsensors based on silica have gained special popularity due to the improvements that they impart on the properties of encapsulated dye, such as photostability[2], good cytocompatibility[3] and flexibility for easy functional modification[4]. To incorporate sensing dye molecules into silica matrix, common ways involve covalent linking, matrix entrapment or charge based immobilization[5]. Covalent linking requires presence of certain functional groups on the dye molecule[1f], which is not possible many of the times. In addition, covalent bond formation can also cause unwanted change in the fluorescence properties of the dye and its sensitivity. Charge based immobilization of positively charged dye molecules is possible due to their interaction with negatively charged silica, however many of the dye molecules are negatively charged and cannot be tethered using covalent bonding. One such particularly useful pH sensitive dye molecule is HPTS (8-Hydroxypyrene-1,3,6-trisulfonic acid trisodium salt), which is pyrene based and offers an extremely sensitive ratiometric pH-sensitive fluorescence emission. Compared to HPTS the other fluorophores like seminaphtharhodafluor (SNARF), (2,7-bis-(2-carboxyethyl)-5-(and-6)-carboxyfluorescein) (BCEF) and fluorescein (FITC) are frequently used in development of various kinds of pH sensors. Fluorescein based indicators are quite common pH-sensitive dyes however they are prone to photoquenching and their fluorescence maxima and pKa can change significantly after chemical substitution. SNARF-1 is a ratiometric dye (pka: 7.5) and is superior to fluorescein due to its higher stability and inertness towards other perturbations. Its bioconjugable forms make it one of the widely used pH probes. However, its fluorescence quantum yield is only around 10% that is a significant limitation. Comparatively, HPTS is one of the most inexpensive pH-sensitive ratiometric dye but compared to naphthoxanthene based dyes like SNARF-1 with cell permeant ester, it has no membrane permeant form. Thus, it is difficult to link HPTS with other molecules or substrates and compromise in pH sensitivity has been observed due to chemical conjugation. HPTS shows an extremely high quantum yield (>75%) and is overly sensitive to near-neutral pH values (pKa: 7.2). In addition to being cell impermeant and non-toxic, pH values of as low as pH 4.4 can be easily measured using HPTS. Large stokes shift is one of the most desired features of any sensor dye and HPTS is one such dye where the Stokes shift is more than 100 nm. Its high polarity makes it highly water-soluble.[6] Despite all its advantages, HPTS is still not fully exploited in the development of nanosensors, where the reason lies in difficulty to tether or immobilize it. Its high negative charge makes it difficult to be used along with negatively charged substrates such as silica and to covalently attach it using its sulphonic acid groups to other substrates which requires harsh chemical treatment and could also potentially compromise with the pH sensitivity. In the past, attempts have been made to use cationic polyelectrolytes to negate the overall negative charge of the silica to make its surface positive enough to attract negatively charged dye molecules, however, a significant change in their fluorescence properties was observed decreasing the overall applicability.[7]

The current HPTS immobilization methodologies on planar as well as optical fibers involve dissolving the HPTS molecules in polymeric matrix like poly(vinyl alcohol),[8] poly(2-hydroxyethyl methacrylate)[9] or polyurethanes[10] and then coat the intended planar surface or fiber with the HPTS containing matrix. Covalent immobilization of the HPTS in polydimethylsiloxane (PDMS)-based amphiphilic conetworks has also been reported however the pH sensitivity is significantly lower than free HPTS molecules.[11]

In general, covalent modification methods have obvious advantages like stronger binding of the molecule for enhanced stability, however on many instances it can seriously perturb the intrinsic fluorescence properties of the molecule of interest. For example, reports on HPTS molecule covalently linked to polymeric matrix such PDMS-poly(2-hydroxyethylacrylate) has shown to result in 58-fold change in fluorescence emission ratio for corresponding pH change from pH 5 to pH 9.[11] The free HPTS molecule on the other hand is much more sensitive and can show ~600 fold change in fluorescence emission ratio during transition from pH 4 to pH 8 (Figure 1A and Table 1).

Non-covalent modifications interfere less with the electronic state of the molecule and are thus less likely to influence its sensitivity. However, there could be problems related to leaching of the molecule over time. In our work we have utilized a more efficient non-covalent immobilization technique to hold the HPTS molecules on silica surface while maintaining appreciable pH sensitivity at the same time. Our proposed approach could provide a better solution where HPTS and similar molecules could be coated without incorporating additional polymers such as hydrogels. In addition, the sensitivity of the molecule could also be optimized easily by manipulating the charges.

Herein, we have investigated different methodologies to immobilize HPTS on silica microparticles (SMPs) and developed a conjugation technique which keeps the HPTS bound to the silica particles with significant retention of its pH sensitivity. Here, the effect of silica particle surface charge was found to be linked with retention of HPTS molecules in addition to its involvement in governing the pH sensitivity. In the end, an overly sensitive HPTS-based pH sensor was developed that can act as a platform in the development of other HPTS-based pH sensors. Besides, the design methodology can be used for immobilization of other similar negatively charged molecules on silica-based substrates.

pH sensing with accuracy becomes especially important when minute fluctuations are to be observed in close vicinity of a cell or tumour structure.[12] In biological systems, pH variation is associated with a functional change in protein and other biomolecular entities.[13] These changes reflect at the systemic and organismic level as an outcome of a disease or malfunction.[14] Due to the importance of pH, its accurate sensing under different conditions is very crucial to understand the underlying activities taking place in an area of interest. Among various techniques of pH sensing, fluorescence-based pH sensing offers several advantages, like noninvasiveness, sensitivity and ability to monitor it at very small scales in a biological system. Therefore, continuous efforts have been made in creating different kinds of fluorescence-based sensors with better biocompatibility, stability, sensitivity. For this purpose, small molecule-based pH sensors like fluorescein, seminaphtharhodafluor (SNARF) and trisodium 8-hydroxypyrene-1,3,6-trisulfonate (HPTS) have been used frequently in their molecular form, however, confining or linking them to nano/microparticles has obvious advantages like higher photostability and better signal to noise ratio etc.[15] Despite these advantages, often fluorescent labelling of nano/microparticles with pH-sensitive small molecules have been known to cause modification in the fluorescence properties of the probes. Sometimes these modifications cause a drastic reduction in pH sensitivity or overall emission intensity or both. These potential problems are thus important to be considered before designing a nano/microparticle-based sensor and choosing the combination of dye molecules and nano/microparticle composition. A very appropriate example of these undesired changes in pH sensitivity could be observed with pyranine, which is also known as Trisodium 8-hydroxypyrene-1,3,6-trisulfonate (HPTS). HPTS is a highly sensitive, inexpensive, water-soluble and membrane-impermeant pH indicator with a pKa of ~7.3 in aqueous buffers. In addition, its large stokes shift, and possibility of ratiometric pH sensing capability makes it special compared to other pH sensitive fluorophores, which either require a reference fluorophore or have a very narrow stokes shift.[16] The ratiometric pH sensing that exists due to pH-dependent shift in its absorption can be realized by exciting HPTS at 405 nm and 450 nm, where both excitations cause emission with maxima at 511 nm and the ratio of emission at respective excitation wavelength can be correlated to pH to create a standard curve.[17]

In the past, the reports on HPTS-based silica nanoparticles for pH sensing exhibited shifted or less sensitive pH response compared to the HPTS molecule dissolved in solution.[7, 18] The reason for the modification was associated with the change in HPTS fluorescence properties caused by the confinement of the dye in the silica matrix or its closeness to the surface that can affect the degree of ionization of the dye against different pH values. Here, we have used physical entrapment and electrostatic interaction to hold the HPTS molecules on the silica substrate instead of a covalent linkage. The reason lies in the fact that covalent binding of HPTS molecules through its sulfonic acid groups is known to decrease the pKa for each bond.[12, 19] Therefore, it can strongly hold the dye molecules but will significantly change the pH sensitivity, which is not desired. In this work, we have investigated unique noncovalent methodologies to incorporate HPTS dye molecules on SMPs and studied the reason behind associated advantages and disadvantages for each of them. The outcome of these studies resulted in an extremely sensitive and stable HPTS-based pH microsensor.

We have investigated four different methodologies immobilizing the HPTS molecules on SMPs (Scheme 1) and we studied the corresponding variation in both the stability and sensitivity of the resulting pH microsensors. We started with the synthesis of highly positively charged SMPs, where HPTS molecules were held onto the surface by electrostatic interactions. These microparticles were named T1-HPTS. In terms of pH sensitivity, T1-HPTS were observed to be most sensitive but highly unstable, as the dye molecules easily wash out in a few washing steps. For each respective pH value, the emission ratio was calculated by dividing the fluorescence emission intensities for excitation wavelengths of 405 and 450 nm (Em: $\lambda_{ex: 405\,nm}$/Em: $\lambda_{ex: 450\,nm}$). The sensitivity of T1-HPTS particles was 13.5 % to that of free dye molecules, indicating minimum possible hindrance caused by SMPs (Figure 1a). The excellent pH sensitivity observed for these sensors can be attributed to the fact that the dye molecules are existing mostly in an electropositive environment with the zeta potential of the T1-HPTS particles as +26.1 mV (Figure 1b). Such condition is known to help in balancing the negative charge of the silica surface that can interfere with the HPTS pH sensitivity. Hence to a large extent, the pH-dependent ionization behavior and the corresponding fluorescence emission change for T1-HPTS was unperturbed.

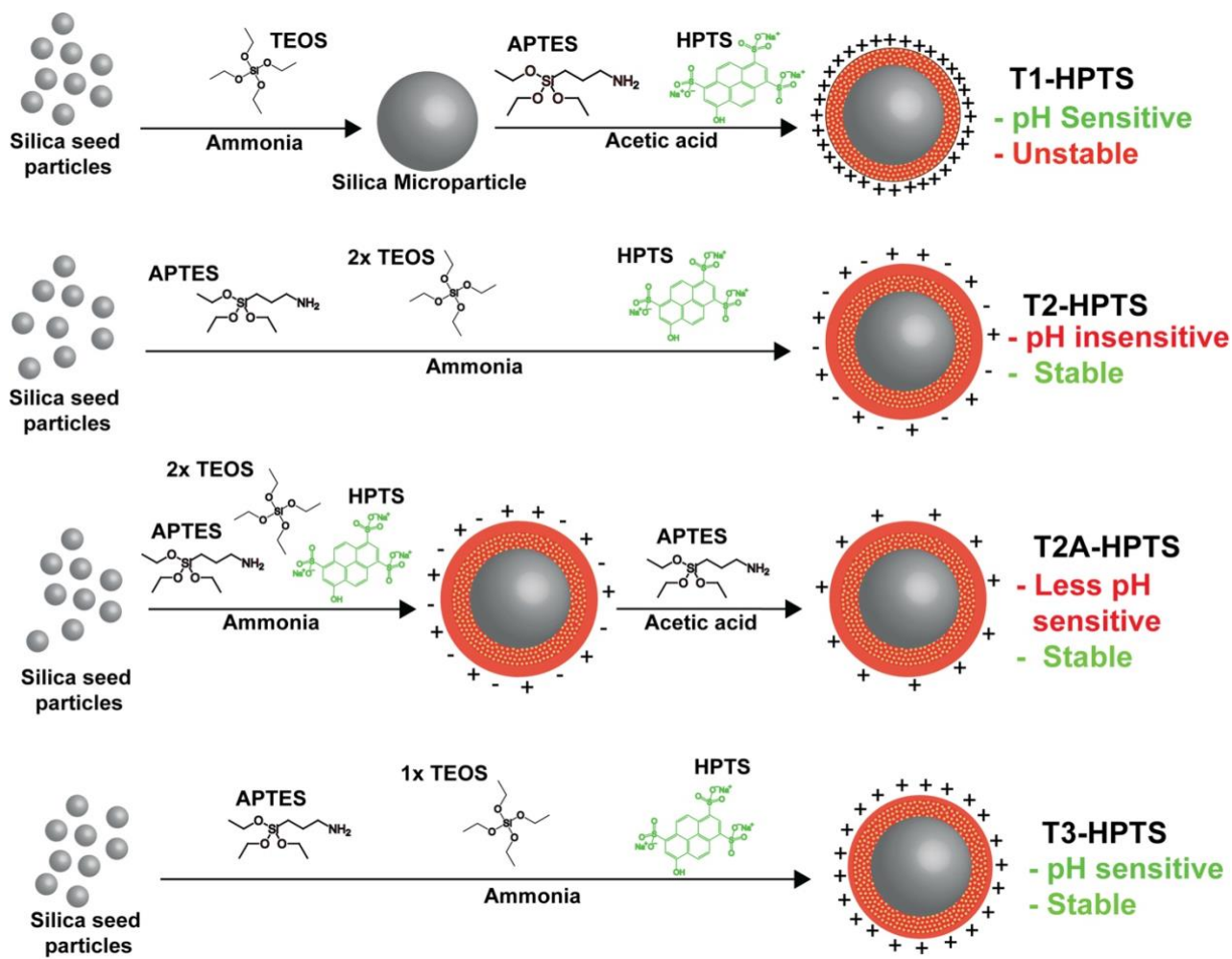

**Scheme 1.** Different methodologies of HPTS immobilization on SMPs, the difference in their pH sensitivity and stability.

Although the sensitivity in case of T1-HPTS was exceptionally good, its stability was an issue. Therefore, we decided to protect the dye molecules by embedding it in a thick layer of the silica shell. This second type of microparticle sensor was named T2-HPTS, where the coating of HPTS was formed around silica seed particles by slowly growing layers of silica using APTES and TEOS as monomers in presence of HPTS. Here, TEOS provided strength to the growing silica layer but brought negative charge to the surface due to hydroxyl groups that can significantly repel negatively charged HPTS molecules. Therefore, to reduce the overall negative charge of TEOS, APTES was used as comonomer, to allow sequestration of negatively charged HPTS molecules during silica microparticle growth. APTES in addition to becoming part of the growing silica shell has a primary amine group, therefore in its ionized form, its positive charge attracts and holds HPTS molecules. The combined use of TEOS and APTES layer polymerization prevented leaching of HPTS molecules and significantly enhanced the stability of T2-HPTS microparticles, however, the T2-HPTS particles showed no clear trend in fluorescence emission change as visible for T1-HPTS (Figure 1a). While T1-HPTS showed an exponential increase in fluorescence emission ratio from 0.36 at pH 8 to 52 at pH 4, T2-HPTS showed a fluctuating emission ratio, which increased from pH 8 to 7 then decreased for pH 6 and again increased for pH 5 and 4.

To understand the reason behind the perturbed pH sensitivity in T2-HPTS, we assessed the zeta potential and found that zeta potential of T2-HPTS was +2.97 mV which was 88.6% less than that for T1-HPTS (+26.1 mV). Thus, it was suspected that the presence of just positive surface charge is not enough and there needs to be a minimum threshold surface charge for the desired pH sensitivity. To test our hypothesis, we decorated the surface of T2-HPTS with additional APTES molecules to make the surface charge more electropositive and tested its effect on the pH sensitivity. As supposed, the pH sensitivity increased significantly after the positive charge enhancement in addition to the irregularity of pH vs fluorescence emission rectified. The modified T2-HPTS has been therefore named as T2A-HPTS, where A represented the presence of additional primary amine groups from APTES. The comparative analysis showed that T1-HPTS was still most electropositive with a zeta potential of about +26.1±5 mV followed by T2A-HPTS and T2-HPTS with zeta potentials as +6.02 ±4 mV and +2.97 ±4 mV, respectively. Here, it is to be emphasized that the stability of T2A-HPTS microparticles was similar to that of T2-HPTS particles, thus the reaction to enhance the positive charge on the particle surface caused no change in stability of the immobilized dye molecules.

Motivated with the attainment of both pH sensitivity as well as stability with T2A-HPTS pH sensors, we decided to further optimize the synthesis protocol to attain best pH sensitivity without losing stability. These microparticles were synthesized similar to that of T2-HPTS, however, the concentration of TEOS is halved. The rationale behind the reduction in TEOS concentration was to increase the overall positive charge on the particle's surface while still leaving enough TEOS to physically lock the HPTS molecules on the surface. This strategy significantly enhanced the pH sensitivity of the resulting microparticles (T3-HPTS). Comparatively, T3-HPTS were observed to be much better than T1-HPTS in terms of stability and more sensitive compared to T2A-HPTS, where the emission ratio changes from 0.36 to 84 for pH change from 8 to 4. Therefore, the percent change in emission ratio with pH change is highest for T3-HPTS (233 times) among stable microparticles (Table 1). As expected, the reason for the enhanced pH sensitivity was also evident from the zeta potential which increased to +11.5±5.2 mV.

Thus, we concluded that charge of the immediate environment of the dye molecule can significantly affect its sensitivity by perturbing the ionization efficiency of the charged dyes as also reported elsewhere.[20] While T1-HPTS was highly positive and showed superb pH sensitivity it was not stable and showed loss of dye molecules, whereas T2-HPTS was very stable but showed poor pH sensitivity. Therefore, protection of the HPTS dye molecule without disturbing its pH sensitivity required a balance of charge on the microparticle surface. While a high positive charge is beneficial for enhancing the pH sensitivity of the HPTS dye on the surface of microparticle, preparing a microparticle with that much charge by reducing the concentration of TEOS during synthesis can significantly affect the strength of shell holding the dye molecules on the microparticle surface. Therefore, it can cause poor dye retention resulting in significant loss of fluorescence from the microparticles, leaving it blank and non-fluorescent. As observed from different forms of sensors prepared in our study, we found that zeta potential of more than +10 mV is enough for keeping the sensitivity equivalent to that provided by zeta potential of +26 mV (for T1-HPTS).

To assess the monodispersity of the pH microsensors, we selected the best two pH sensor types. i.e T2A-HPTS and T3-HPTS. and analyzed their morphology by means of scanning electron microscopy (Figure 3). The analysis showed appreciably monodispersed spherical microparticles with a diameter of ~962±78 nm and ~790±31 nm for T2A-HPTS and T3-HPTS, respectively. Notably, the microparticles due to their large size compared to nanoparticles have a unique advantage in fluorescence microscopy, that they can be resolved very easily to observe each microparticle individually. The larger size also brings with it a lower surface area to volume ratio that reduces the probability of aggregation in an experimental setup and during long time storage. Additionally, SMPs have been observed to be more cytocompatible as compared to their nanoparticle counterparts that have higher surface energies that make them more interactive against cellular organelles and biomolecules.

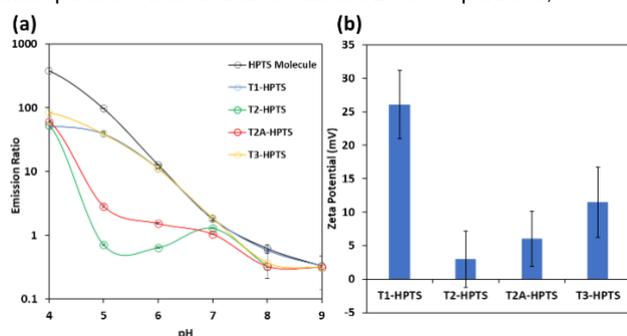

**Figure 1. (a)** Comparison of pH sensitivity of HPTS based fluorescent microsensors using different strategies. **(b)** surface charge of different HPTS based microparticles due to the difference in synthesis methodology.

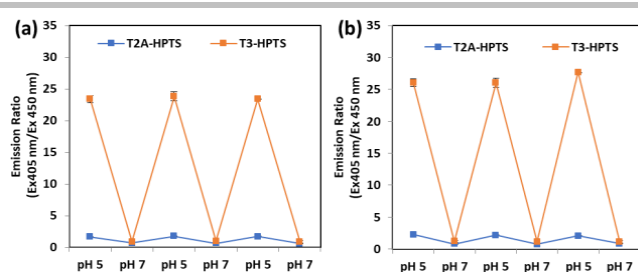

**Figure 2.** **(a)** Reversibility of T2A-HPTS and T3-HPTS evaluated conducting sequences of 3 pH switches between pH 5 and pH 7 (n=3). **(b)** Ageing of HPTS-aminated and HPTS sensors evaluated repeating sequences of 3 pH switches between pH 5 and pH 7 (n=3) after 7 days on the samples adopted to evaluate the reversibility.

T2A-HPTS and T3-HPTS sensors reversibility and ageing were also investigated (Figure 2). Reversibility was evaluated by a series of three pH switches between pH 7 and pH 5. At least 3 samples for each type of sensor were tested. T3-HPTS sensors appeared more sensitive compared to T2A-HPTS showing higher ratios values, with average emission ratios equals to 21.86 and 0.13 for pH 5 and pH 7, respectively (Figure 2a). As regard the sensors ageing for T2A-HPTS and T3-HPTS, series of three switches between pH 7 and pH 5 were repeated after 7 days on the same samples employed to evaluate reversibility to assess their sensing ability despite the ageing. Again, all observed data resulted statistically significant accordingly to ANOVA with single factor ($\alpha = 0.05$) with very limited standard deviation and variance values on both T2A-HPTS and T3-HPTS, thus proving robustness and response reliability of both systems.

The two graphs can be considered directly comparable as the back calculated pH values from the emission ratios would be very similar in both the cases. This robustness is observed due to the fact that the change in the emission ratio with pH change is exponential as can be observed in Figure 1a. For T3-HPTS the observed change in emission ratio is 233 folds when the pH is changed from pH 8 to pH 4. For this large change in emission ratio, even a minuscule fluctuation in pH due to instrumental error of pH meter, can result in an observable change in emission ratio.

Reversibility studies with T1-HPTS and T2-HPTS were not conducted as T1-HPTS particles were not very stable causing significant amount of dye molecule being lost during washing as well as pH balance steps. Therefore, it was unreliable to test the pH reversibility for T1-HPTS particles. T2-HPTS was excluded from reversibility study because its range of pH sensitivity was not following a trend as shown in Figure 1a.

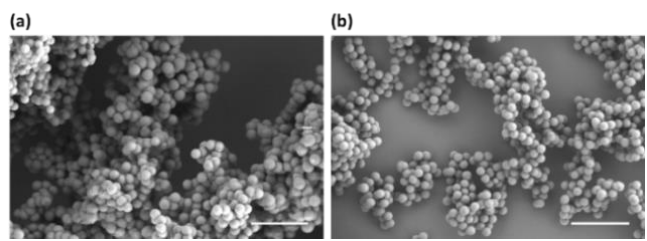

**Figure 3.** Scanning electron micrographs of **(a)** T2A-HPTS and **(b)** T3-HPTS microparticles with diameter of ~962 nm and ~790 nm, respectively (scale bars: 5 μm).

Motivated by the encouraging results for T3-HPTS, we also tried to assess the pH sensitivity of T3-HPTS by using confocal laser scanning microscopy (CLSM), where a significant change in emission intensity was observed when pH was changed from pH 8 to pH 4 (Figure 4). The panels represent sensor particles imaged at different pH values. Using images acquired under different excitation wavelength, merged and ratiometric images were generated and color coded for showing ratiometric pixel values. Ratiometric images were obtained by dividing the intensity of individual pixels of images acquired using 405 nm and 458 nm excitation and creating a third image with the obtained pixel ratios.

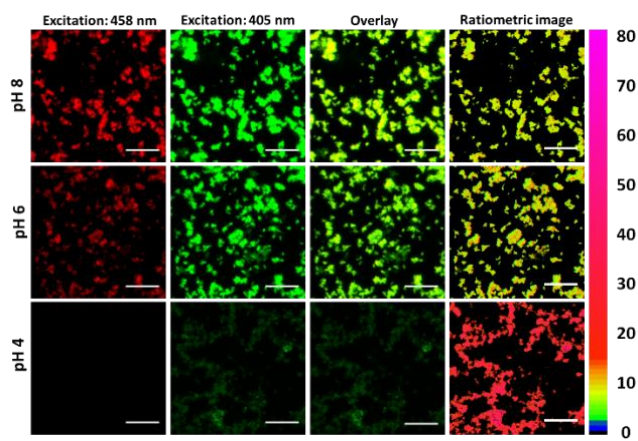

**Figure 4.** Fluorescence confocal microscopy images showing pH-dependent fluorescence emission of T3-HPTS in independent emission channels, overlay and ratiometric ($\lambda_{ex:405\ nm}$ / $\lambda_{ex:458\ nm}$). The emission was detected in the range 474nm - 570nm. Scale bars: 20 μm.

Owing to the cytotoxicity caused by particle-based sensors, viability of HCT116 cells were assessed after their incubation with T3-HPTS sensors. To this aim, we incubated colon cancer cell line HCT116 with T3-HPTS sensors. Compared to the positive control (untreated cells), the cells incubated with the T3-HPTS sensors showed equal viability at different incubation time points as indicated in figure S1. The statistical analysis proved that there was no significant difference between viability of treated and untreated cells. This nontoxic behavior is most likely due to bio-inert behaviour of silica[21] and the micrometer size of these particles that significantly reduce the surface energy that is primary cause of interaction of the nanoparticles with biomolecules [22] and cell organelles.[23]

Next, the pH sensitivity and cytocompatibity analyses of T3-HPTS, we tested the uptake of T3-HPTS sensors by mammalian cells. To this aim, we incubated colon cancer cell line HCT116 with T3-HPTS pH sensors for 24 hours followed by fluorescence imaging. At time 0 hours, the HPTS particles were mainly distributed in the cell medium (neutral pH) and, in agreement with the HPTS pH-sensitivity, showed fluorescence emission from both 405 and 458 nm excitation (Figure S2). At 24 hours, the majority of T3-HPTS pH sensors were detected in close proximity or inside the cells and showed a diminished fluorescence emission at 458 nm excitation. In agreement with other studies based on different pH probes[1a, b, 24], the change of fluorescence of HPTS might reflect the intracellular localization of the T3-HPTS particles in an acidic environment[25] (e.g., endosomes or lysosomes), causing the observed decrease of emission from excitation at 458 nm. Additional experiments by electron microscopy aimed at deeply investigating the cell uptake of the T3-HPTS sensors and their intracellular localization are currently underway in our laboratories.

Table 1 shows the main properties of the different types of HPTS-based sensors. The relative pH sensitivity for each type of sensor was calculated using the sensitivity of molecular HPTS as a reference. This comparison showed that T1-HPTS had the mean sensitivity of 13.5% with a pKa of about 6.53 which was less than that of free HPTS. T2-HPTS on the other hand showed similar mean sensitivity but large standard deviation owing to its irregular pH vs emission ratio behavior. T2A-HPTS showed pH sensitivity of almost 16%, which was higher than T2-HPTS, but less than the sensitivity of T3-HPTS (22%). As can be observed, for all the sensors the pKa is less than that of free HPTS pKa which is ~7.25.[11, 26] However, among all the analyzed systems, the change in pKa is minimum for T3-HPTS (pKa: 6.97). Thus, the final optimized methodology for pH sensor development is suitable for sensitivity, stability as well as maintaining the intrinsic behavior of the fluorophore.

**Table 1.** Comparison of sensitivity, stability and pKa for different HPTS-based sensors developed.

| Sensor types | Sensitivity to pH | Emission ratio | | % change w.r.t. HPTS | pKa | Stability |
|---|---|---|---|---|---|---|
| | | pH 8 | pH 4 | | | |
| HPTS (ref) | Extremely high | 0.61 | 380.15 | 100 | 7.25 ±0.17 | -- |
| T1-HPTS | Very High | 0.58 | 52.05 | 13.5 | 6.53 ±0.12 | Low |
| T2-HPTS | Low | 0.32 | 53.55 | 14 | 5.76 ±1.21 | High |
| T2A-HPTS | Very high | 0.32 | 60.92 | 15.9 | 6.15 ±0.59 | High |
| T3-HPTS | Very High | 0.36 | 84.04 | 22 | 6.97 ±0.19 | High |

In conclusion, we have devised ways to successfully use negatively charged pH-sensitive HPTS molecule with SMPs without compromising its pH sensitivity. These sensors could be used in sensing pH *in vitro* as well as *in vivo* systems due to the cytocompatibility of HPTS molecules and SMPs. These HPTS based pH microsensors due to their stability and sensitivity could potentially be used for endosomal pH tracking as well as extracellular pH assessment in *in vitro* studies involving spheroids and organoids. In future we will be exploring their utility in these directions.[27]

The extreme change in emission ratio with even a small change in pH will help in investigating very subtle changes in pH. The methodology used here can be employed to immobilize various other anionic dye molecules and even anionic drugs on the silica surface. New sensing platforms could be developed using this methodology to build sensitive optical fiber probes that have potential application in *in vivo* applications like measurement of tumor tissue acidity. In similar fashion planar surfaces could be layered with electrostatically incompatible charged dye molecules using the current methodology. This would help in development of miniature sensor surfaces integrated into microfluidic devices where glass substrate could be modified with dye molecules which are difficult to tether.[27c]

As emphasized previously, the balance by tailoring surface charge and protecting the molecule against leaching. This would ultimately help in realizing more sensitive and robust sensing devices and platforms.

## Acknowledgements

The authors are grateful to the European Research Council (ERC) under the European Union's Horizon 2020 research and innovation programme (grant agreement No. 759959, INTERCELLMED), the AIRC under MFAG2019 (ID. 22902) and Tecnopolo per la medicina di precisione - Regione Puglia (project number: B84I18000540002).

**Keywords:** pH sensors, HPTS, Microparticles, Fluorescence, Ratiometric sensing.

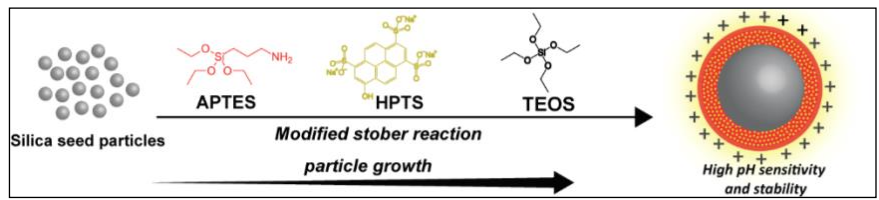

**TOC text**

In a one-step synthesis, silica microparticles with efficiently immobilized Pyranine (HPTS) dye resulted in the development of overly sensitive fluorescent ratiometric pH sensors. The methodology to immobilize dye molecules on oppositely charged silica microparticles by utilizing a positively charged silica shell preserves HPTS pH sensitivity, and also results in enhanced stability of the sensor particles. Effect of microparticle surface charge manipulation for enhancing pH sensitivity is explored without perturbing the stability and cytocompatibility.